\documentstyle[aps, twocolumn]{revtex}

\begin{document}
\author{P. Mendels$^{1}$, A. Keren$^{1,2}$, L. Limot$^{1}$, M. Mekata$^{3}$, G.
Collin$^{4}$ and M. Horvati\'{c}$^{5}$}
\title{Ga NMR study of the local susceptibility in SrCr$_{8}$Ga$_{4}$O$_{19}$:
pseudogap and paramagnetic defects}
\address{$^{1}$Laboratoire de Physique des Solides, UMR 8502, Universit\'{e} Paris-Sud, 91405 Orsay, France.\\
$^{2}$Department of Physics, Technion, Israel Institute of Technology, Ha\"{i}fa 32000, Isra\"{e}l\\
$^{3}$Department of Applied Physics, Fukui University, Fukui 910, Japan\\
$^{4}$Laboratoire L\'{e}on Brillouin, CE Saclay, CEA-CNRS, 91191 Gif-sur-Yvette, France\\
$^{5}$ Grenoble High Magnetic Field Lab., CNRS\ and MPI-FKF, BP 166, 38042 Grenoble Cedex 9, France.}
\maketitle

\begin{abstract}
We present the first Ga(4f) NMR study of the Cr susceptibility in the
archetype of Kagom\'{e} based frustrated antiferromagnets, SrCr$_{8}$Ga$_{4}$%
O$_{19}$. Our major finding is that the susceptibility of the frustrated
lattice goes through a maximum around 50 K. Our data also supports the
existence of\ paramagnetic ``clusters'' of spins, responsible for the Curie
behavior observed in the macroscopic susceptibility at low $T$. These
results set novel features for the constantly debated physics of
geometrically frustrated magnets.\ 
\end{abstract}

\date{\today}

The interest in triangular-based antiferromagnets (AF) was raised long ago
by Anderson's suggestion for a resonating valence bond state (R.V.B) as an
open alternative to the classical N\'{e}el state \cite{Anderson}. After a
decade of extensive work on 2D geometrically frustrated AF, there is now a
growing theoretical consensus that the $S=\frac{1}{2}$ Kagom\'{e} Heisenberg
AF is a very promising candidate for such an RVB state at low $T$. Instead
of sharing bonds, as in a triangular lattice, the frustrated triangles share
sites with a smaller coordinance. From a classical point of view, \cite
{Classtheory,Reimers} this generates a very high degeneracy of the ground
state, which translates into a huge density of low energy excitations, an
absence of long range order at $T=0$ and a very short magnetic correlation
length characteristic of a so-called ''spin liquid state''.\ Quantum
approaches for $S=1/2$ spins suggest that a singlet ground state is
energetically favoured \cite{L'Huillier,Mila}.

Only some of these theoretical features were demonstrated experimentally on
the archetypal system SrCr$_{9p}$Ga$_{12-9p}$O$_{19}$ (SCGO:$p$)
Kagom\'{e}-based AF. Susceptibility measurements \cite{Martinez,Ramirez}
showed the strong AF character of the interactions through the occurrence of
a Curie-Weiss (CW) temperature as high as $\theta \,\simeq \,500-600$ K (for 
$p\simeq 0.9$). The extension of the CW law well below $\theta $ and the
smallness of the spin glass like ordering detected in susceptibility only at
low $T\ll \theta ,$ ($T_{g}=3-5$ K) are taken as convincing signatures of
the high frustration. In addition, neutron data \cite{Neutron} and low $T$ $%
\mu $SR studies \cite{muSR} altogether prove that the freezing is quite
marginal and involves only a 20-30\% fraction of the moment(s). The magnetic
correlation length is also found of the order of the inter-Cr spacing.

However, due to a Curie upturn which dominates the macroscopic
susceptibility, $\chi _{macro}$, below 60 K\cite{Martinez,Ramirez} there is 
{\it no} experimental determination of the $T\ll \theta $ Kagom\'{e}-based
lattice susceptibility. The Curie upturn received many interpretations still
highly debated, encompassing magnetic disorder due to non magnetic
substitutions in the Kagom\'{e} plane \cite{Daruka,Berlinsky} or original
geometrically frustrated spin glass like order \cite{Ramirez}.

In this Letter, we present the first NMR study of {\it local} susceptibility
in SCGO, where we can discriminate between the susceptibility associated
with the Kagom\'{e} based frustrated magnetism, $\chi _{frust}$ \cite
{Kago/pyro}, and the purely paramagnetic susceptibility, $\chi _{def}$, most
likely induced by Ga/Cr substitutions (defects). Strikingly, we find a
gap-like downturn of $\chi _{frust}$ below $T$ as high as 50 K. This sets a
novel energy scale, between $\theta $ and $T_{g}$ for the relevant physics
of SCGO.

From $\chi _{macro}$\ measurements, the SCGO:$p=0.9$\ sample was found to be
typical for such a Cr content \cite{Martinez}, with $\theta =560$ K and a
high-$T$\ Curie constant, $C=2.2$ emu/mol. An extended NMR\ analysis of the
spectrum was presented in a previous paper \cite{GaNMRKeren}. There, it was
clearly established that 3 different Ga NMR sites could be resolved, Ga($4e$%
), Ga($4f$) and Ga substituted on Cr sites, Ga($sub$).\ We focus here on the
Ga($4f$) site, whose line shift probes both the Cr($12k$) Kagom\'{e} and Cr($%
2a$) triangular planes susceptibility through the neighboring O\cite
{GaNMRKeren}. Interestingly, these sites form the so-called pyrochlore slab 
\cite{Lee,Kago/pyro} (fig.1).

Below $T=200$ K the $^{71}$Ga spectra were recorded by sweeping the field,
using a conventional $\pi /2$ - $\pi $ spin echo sequence to detect the $%
^{71}$Ga($4f$) and $^{69}$Ga($4f$) nuclear transitions. Compared to Ref. 
\cite{GaNMRKeren}, a much lower NMR frequency $\nu _{0}\simeq 40$ MHz was
used to ensure a better suppression of the Ga ($4e$) contribution. This is
achieved by taking advantage of the quadrupole line broadening ($\simeq \nu
_{Q}^{2}/\nu _{0}$) which is strongly different for the two sites which have
a different local charge environment ($^{71}\nu _{Q}(4f,4e)\simeq 2.9,20.5$
MHz). For $220<T<450$ K, the NMR specta were taken by sweeping the frequency
in a fixed 7.5 T field.

A typical set of the field sweep spectra, recorded around $3$T, is reported
in Fig. 2. The high $T$ part (upper panel), show that upon cooling the NMR
lines shift to the right (lower $H$), without any appreciable broadening. In
contrast, the low $T$ part (lower panel), shows that upon further cooling
the lines shift to the left (higher $H$), and broaden. This crossover
(taking place at $\sim 50$~K) in the $T$ - dependence of the shift,
reflecting the local magnetic susceptibility, is the major finding of this
letter.

Another feature seen in Fig. 2 is a wipe-out of the intensity due to fast
nuclear relaxation when the dynamics of electronic spins slows down. This is
common in various systems ranging from spin glasses to AF correlated systems 
\cite{AlloulSG,Mendels-Imai}. From Ref.\cite{AlloulSG} and the $\mu $SR
relaxation data reported on the same compound \cite{muSR,Kerenmusr}, one
would naively expect this effect to occur in SCGO only in the vicinity of $%
T_{g}$. Carefully measured integrated intensity of the $^{69,71}$Ga spectra
shows no variation above 15 K (fig.3), clearly demonstrating that our data
very reliably reflects the behavior of all the electronic spins in the
system, including the $T$-range 15-50 K where the shift direction changes.
On the contrary, it is worth noticing that below $T=10$ K$\sim 3T_{g}$, more
than 50\% of the sites are wiped out of our experimental window, indicating%
{\it \ }the occurrence of an inhomogeneous dynamics of the spin system \cite
{AlloulSG} in an unusually high{\it \ }$T$- range as compared to $T_{g}$.

The shift $K$ and the width $\Delta H$ are extracted from the NMR\ line at
all $T$. $K$ is related to the average field at the Ga($4f$) site, and
directly probes the (homogeneous) susceptibility of the Cr$(12k)$ and Cr($2a$%
) ions. $\Delta H$ originates from a distribution of internal fields on the
nuclear Ga($4f$) site, which is naturally associated with an {\it %
inhomogeneous} susceptibility of the Cr spin system. For $T>120$ K, the line
broadening is small enough that, for practical purpose, we extract $K$ from
the shift of the line edges. This method is not adequate at lower $T$. Below
120 K, the line is symmetrically broadened, and to deduce $K$, we used
either the centre of gravity or a partial Gaussian fit of the Ga(4f)
contribution. Independently of the type of analysis,{\it \ }$K${\it \ }was
found to{\it \ decrease} at low{\it \ }$T$. $\Delta H$ was extracted from
Gaussian fits\cite{Ga(4e)foot}.

First we discuss the temperature dependence of $K$ which is presented in
fig. 4, where we also include results taken at various applied fields. From
the high $T$-data (inset), we can extract a N\'{e}el temperature $\theta
_{NMR}\simeq 470$ K of the same order as $\theta _{macro}=560$ K. This
confirms that $K$ reflects the physics of the frustrated unit. As mentioned
before, $K$ first increases with decreasing $T$ down to $50$ K, but below, $%
K $ {\it flattens} and {\it even decreases} by $20$\%. The sharp contrast
between the temperature dependence of $K$ and $\chi _{macro}$,\ below $50$
K, is emphasized by the dashed arrow in the figure. It reveals that{\it \ 2
different types of Cr} have to be considered. In other words, our shift data
rule out models which attempt to associate the low-$T$ macroscopic
susceptibility only with a generic - therefore homogeneous- property of the
frustrated lattice. Further investigations, to be detailed elsewhere \cite
{LaurentNMR}, clearly confirm that the shift variation reported for this
sample is an intrinsic feature of the frustrated network as it is very
little dependent on the Cr/Ga substitution (at variance with the width).

Next we discuss the variation of $\Delta H$ at low $T$. The results, taken
for various frequencies, are summarized in fig.5. At low $T$, the broadening
scales remarkably with the applied field for both isotopes. This clearly
confirms the magnetic origin of the width at low $T$\cite{GaNMRKeren}. A
Curie-like behavior is found for $\Delta H(T)$, as shown by the solid line.
We therefore plot in the inset $\Delta H/H_{0}$ versus $\chi _{macro}$
measured for the same $H_{0}\sim $3 Tesla field, using $T$ as an implicit
parameter. The linearity of the relationship between $\Delta H$ and $\chi
_{macro}$ strongly suggests that{\it \ }the Curie upturn, which dominates $%
\chi _{macro}$\ at low $T$, and the linewidth have a common origin of {\it %
inhomogeneous} magnetism. The deviation between $\chi _{macro}$ and $K$
below $T=50$~K is also straightforwardly explained by this viewpoint.

In summary, our NMR results are consistent with a picture where $\chi
_{macro}$ is a sum of two distinguished components $\chi _{frust}$ and $\chi
_{def}$. $\chi _{frust}$ is the homogeneous susceptibility reflected in $K$
and representing the physics of the kagom\'{e}-based lattice. It has a
Curie-Weiss like behavior at high $T$ and displays a crossover at $50$ K to
a pseudogap behavior. $\chi _{def}$ is the inhomogeneous contribution to the
susceptibility reflected in $\Delta H$ and originating from defects of the
frustrated block. This component has a pure Curie low-$T$ contribution and
it dominates $\chi _{macro}$ at $T\rightarrow T_{g}^{+}$.

We now turn to discuss our experimental results in light of existing
theories. Susceptibility of the pure Kagom\'{e} network has been numerically
simulated using various models. In many of them, such as {\it e.g.} the case
of singlets formation \cite{L'Huillier}, a gap $\Delta $ appears in $\chi
_{frust}$ and $\chi _{frust}\sim e^{-\Delta /T}$ at low $T$. Using for $%
\Delta $ the temperature $T_{\max }=50$ K where $K$ peaks, one would expect
a much sharper decrease of $\chi _{frust}$ than the 20\% decrease observed
at $20$ K. The discrepancy between the theoretically expected and measured
decrease of $K$ might be solved using a more realistic model of pyrochlore
slab \cite{Berlinsky}. In this model spins from the triangular Cr($2a$)
layer combine with the Kagom\'{e} Cr($12k$) to generate a basic unit with an
uncompensated moment. This moment is expected to add a $1/T$ homogeneous
contribution to $\chi _{frust}$, which should weaken the drop of $K$ below $%
\Delta $. Whether such a term somewhat counterbalances the effect of the gap
on the measured $K$ is still speculative as, unfortunately, an experimental
confirmation is prevented by the loss of NMR intensity below 15~K.
Therefore, we cannot definitely conclude on the full opening of a gap at the
present stage, hence the name pseudogap.

Regardless of the nature of the gap, the value of $T_{\max }$ is very
surprising. In most models $T_{\max }<0.1J$ where $J$ $\sim $ $100K$ \cite
{Lee} is the exchange interaction. Here $T_{\max }\sim J/2$, which is much
bigger than expected and obviously further theories are required to explain
in detail our results. To our knowledge, only a chiral model features a peak
in $\chi $ at $T$ as high as $0.4\,J$ \cite{Reimers}. From the absence of
neutron signature, one does not expect any real magnetic order to occur
around 50K. Our shift data would rather indicate an increase of the magnetic
correlations peaked at the chiral wave-vector. This scenario resembles the
case of the pseudogap in High $T_{c}$ cuprates.\ An even simpler
interpretation to the high value of $T_{\max }$ relies on the fact that for
any low dimensional AF correlated system, one expects the susceptibility to
decrease at low $T$. The crossover usually occurs in non-frustrated $2D$ AF
for $T\sim \theta $\cite{deJongh}, however, because of frustration, it could
occur at lower temperatures, here 10 times smaller than $\theta $. The ratio 
$\theta /T_{\max }$ might, finally, prove to be a better characterization of
the degree of frustration than using $\theta $ $/T_{g}$ since the origin of
the spin glass freezing might be associated with defects, as discussed below.

We now propose an interpretation for the origin of the line width in the
light of the model developed in \cite{Berlinsky}. There, the\ origin of the $%
1/T$ paramagnetic behavior of $\chi _{macro}$ is assigned to the existence
of triangles of the Kagom\'{e} lattice non fully occupied by Cr$^{3+}$
moments. The substitution of two adjacent Cr$^{3+}$ sites by Ga seems
necessary to generate a paramagnetic-like ``defect'' at low $T$. {\it A
priori }such a paramagnetic defect could lead to a well defined feature in
the spectrum and a broadening, depending on the response of the electronic
spin system to this defect. The number of\ Ga sites directly coupled to
these $\sim (1-p)^{2}=1\%$ triangles is small and the corresponding signal
is thus likely unobservable. On the contrary, a staggered response (sign
oscillation of the field generated by the defect as a function of distance)
over few lattice constants, is expected to lead to a symmetric line
broadening of the full line. This phenomenon is observed in a large number
of systems such as High $T_{c}$ cuprates, or 1D spin chains and ladders,
where a similar low-$T$ increase of the NMR linewidth \cite{sitesubs} is
reported. Therefore, we conclude that the defects in SCGO must be coupled to
the surrounding correlated spins, in agreement with the idea of Ref.~\cite
{Berlinsky}\cite{darukabis}.

For a quantitative analysis of the low-$T$ contribution of Ga/Cr
substitutions to $\chi _{macro}$, one needs to subtract the contribution
from an ideal pure sample, unfortunately not stable. Nevertheless, we use a
simple (and consistent) viewpoint where $\chi _{macro}$ is dominated by the 
{\it substitution} defects at low $T$. From the value of the low temperature
Curie constant $C_{LT}${\bf \ }deduced from $\chi _{macro}$, we can deduce
the value of the effective moment associated with one defect, $\mu _{eff}$,
provided the number of defects, $N_{defect}$ , is known. We follow \cite
{Berlinsky} and write $N_{defect}/N_{Cr}=3/2\,(1-p)^{2}$, where $N_{Cr}$ is
the total number of Cr. From $C_{LT}=N_{defect}\mu _{eff}^{2}/3k_{B}=0.03$
emu/mol \cite{CLTvalue}, we find $\mu _{eff}($defect$)\sim 4\,\mu _{B}$,
typical of a spin $3/2$. This reminds a similar case for $S=1/2$ AF cuprates
where the absence of spin in the square 2D network generates a staggered
damped response of the surrounding spins with a total moment corresponding
to a spin 1/2\cite{dagotto}. The overall consistency with the model of Ref.~ 
\cite{Berlinsky} is encouraging, but, of course, more NMR and susceptibility
experiments are needed for other low substitution rates, in order to further
check the quadratic concentration dependence of $N_{defect}$.

In conclusion, we have demonstrated that the intrinsic Kagom\'{e}/pyrochlore
slab susceptibility displays a broad maximum around $T\sim J/2$. For $T<20$%
K, our data suggest that the macroscopic susceptibility is dominated by the
contribution from defects which remain coupled to the frustrated network\cite
{Berlinsky}. Finally, the occurrence of a slowing down of spin fluctuations
is clearly evidenced below 15K. Our results definitely set new constraints
on the theoretical models and are stimulating for other NMR studies in the
broad class of frustrated systems.

We acknowledge Y.J. Uemura who suggested this work and C. L'Huillier, F.
Mila, H. Alloul, J. Bobroff for fruitful discussions.

\begin{figure}[tbp]
\caption{Typical $^{71}$Ga($4f$) field sweep spectra obtained for $\protect%
\nu _{0}=40.454$ MHz, plotted versus $(H_{0}-H)/H$ ($H_{0}$ is the
non-shifted value of the resonance field and $H$ is the applied field).
Upper panel: $T>100K$, expanded scale. Bottom panel: $T<50$ K:\ the arrow
indicates approximately the position of the line expected at 25 K in a
one-component model of $\protect\chi _{macro}$ (with $K$ and $\protect\chi %
_{macro}$ scaled at high $T$).}
\label{fig.2}
\end{figure}
\begin{figure}[tbp]
\caption{Number of Ga(4f) sites detected by NMR. The $^{69}$Ga estimate is
more accurate as the ratio of the (4$e$) contribution (quadrupolar
broadening) to the ($4f$) one (magnetic broadening) is 2.5 smaller than for $%
^{71}$Ga. $T_{2}$ corrections were found quite small and do not affect the
estimates below 50 K.}
\label{fig. 3}
\end{figure}

\begin{figure}[tbp]
\caption{$K$ versus $T$ down to 10 K, for various fields / frequencies.
Minor second-order quadrupole corrections have been performed. The spread in
the low-$T$ values taken in various conditions is due to the sizeable line
broadening, see fig.2. The dashed line figures $K$ variation expected from $%
\protect\chi _{macro}$ within a one-component model. Inset:$1/K$ versus $T$.
The straight line extrapolation to $1/K=0$ yields $\protect\theta _{NMR}$.}
\label{fig. 4}
\end{figure}

\begin{figure}[tbp]
\caption{Relative full width at half maximum plotted versus $T$ for the 2
isotopes at various fields. At high $T$, $\Delta H$ is dominated by $T$-
independent quadrupole effects, more prominent for $^{69}$Ga and low fields.
Inset: Plot of $\Delta H/H_{0}$ versus $\protect\chi _{macro}$ in the same $%
T $-range. }
\label{fig. 5}
\end{figure}


\begin{references}
\bibitem{Kago/pyro}  Although SCGO is considered as archetype of Kagom\'{e}
systems, it is not clear whether its physics is purely governed by AF Cr ($%
12k$) Kagom\'{e} planes or is affected by the substantial transverse
coupling between two adjacent Kagom\'{e} layers through Cr($2a$). Treating
all the Cr-Cr bonds on the same footing yields a pyrochlore slab as the
basic unit for any model.
\bibitem{Ga(4e)foot}  The broad Ga(4e) contribution, which appears as a
pedestal (fig.2, bottom) and becomes more prominent below $50$ K, is taken
into account in our fits of $\Delta H$.
\bibitem{darukabis}  A similar picture of paramagnetic moments was
conjectured by \cite{Daruka}, based on a 2-component analysis of the
susceptibility of various frustrated magnets. The moments are there
uncorrelated to other Cr spins, then purely local. Our results show that
this is not the case.
\end{references}
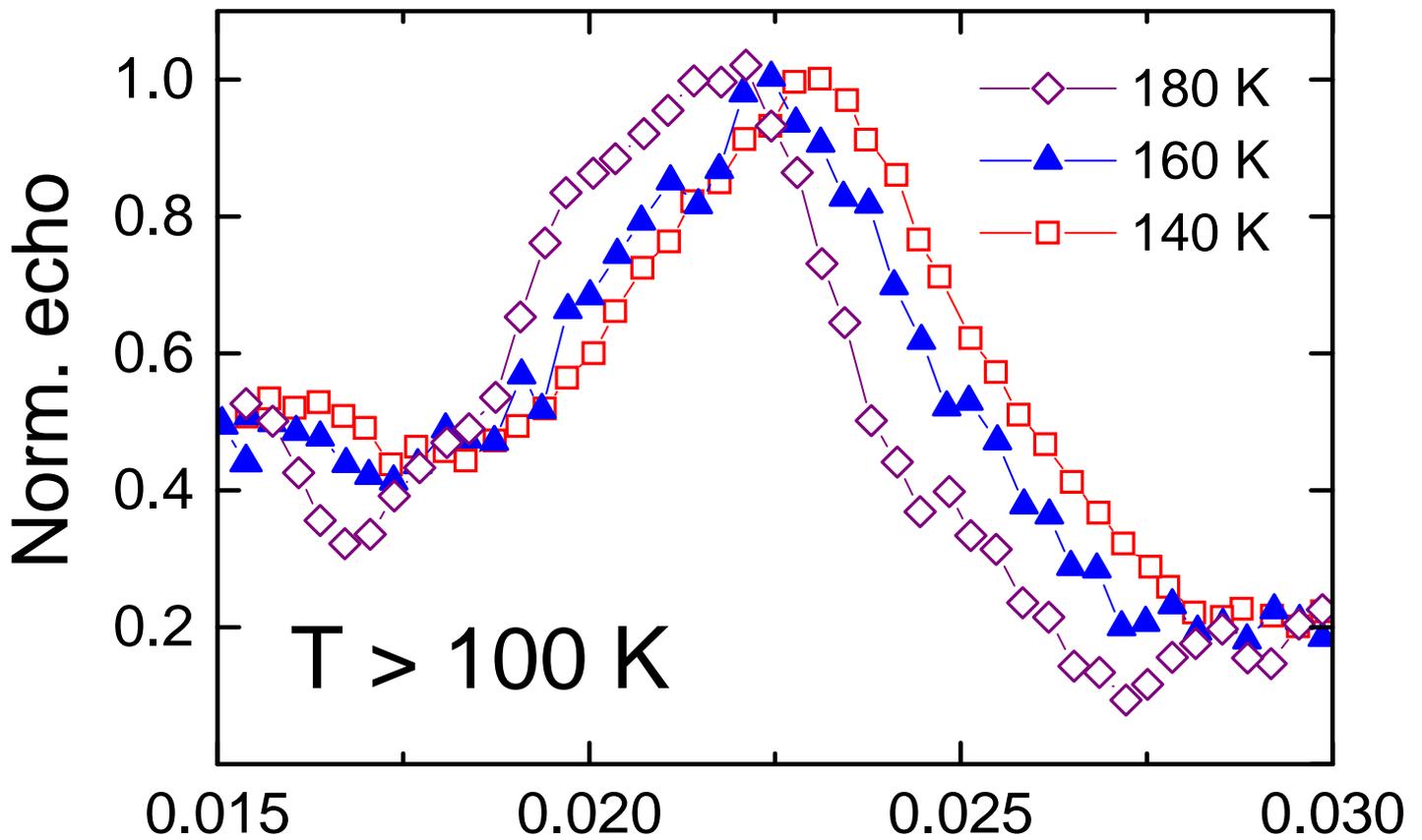
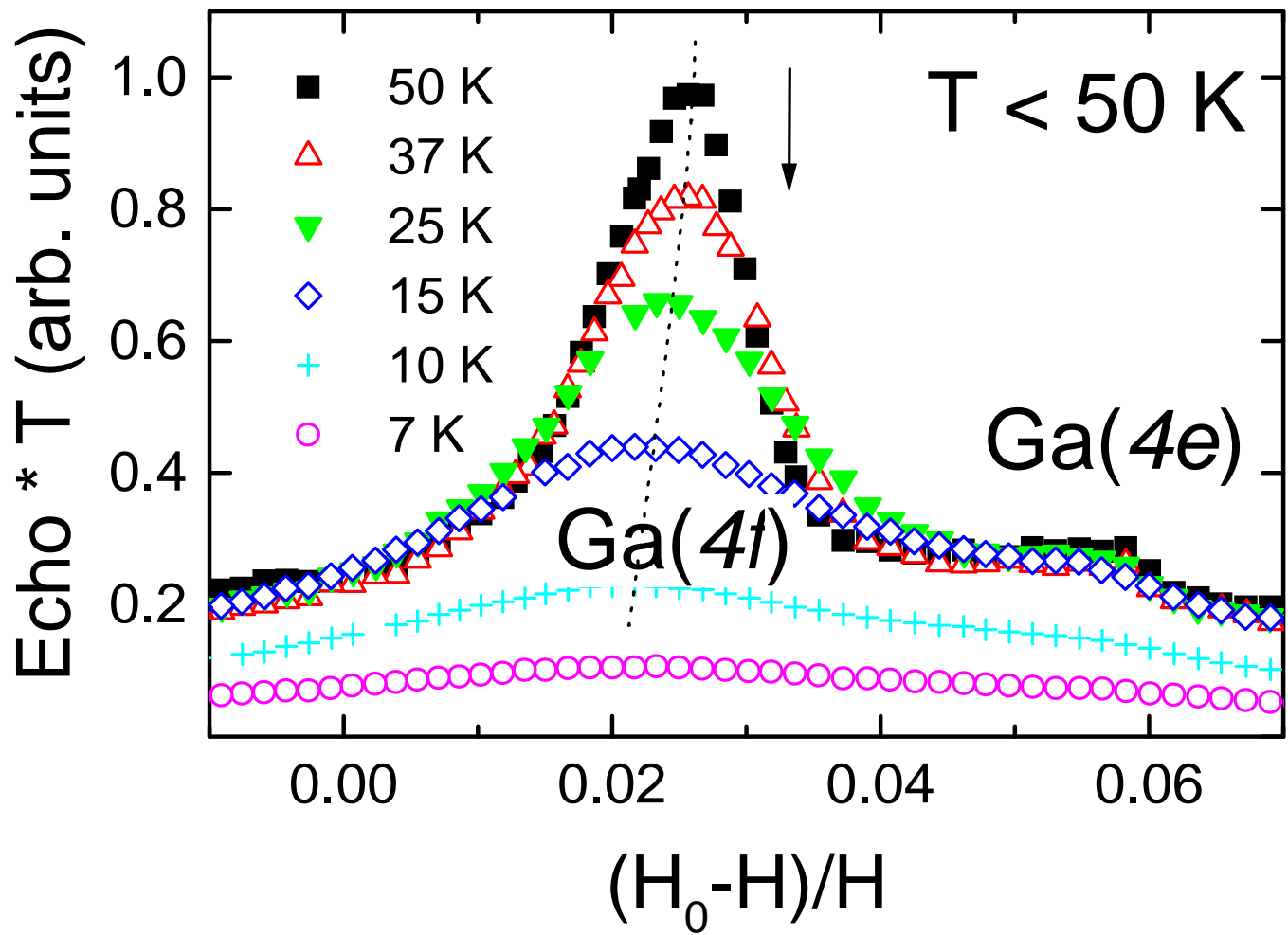
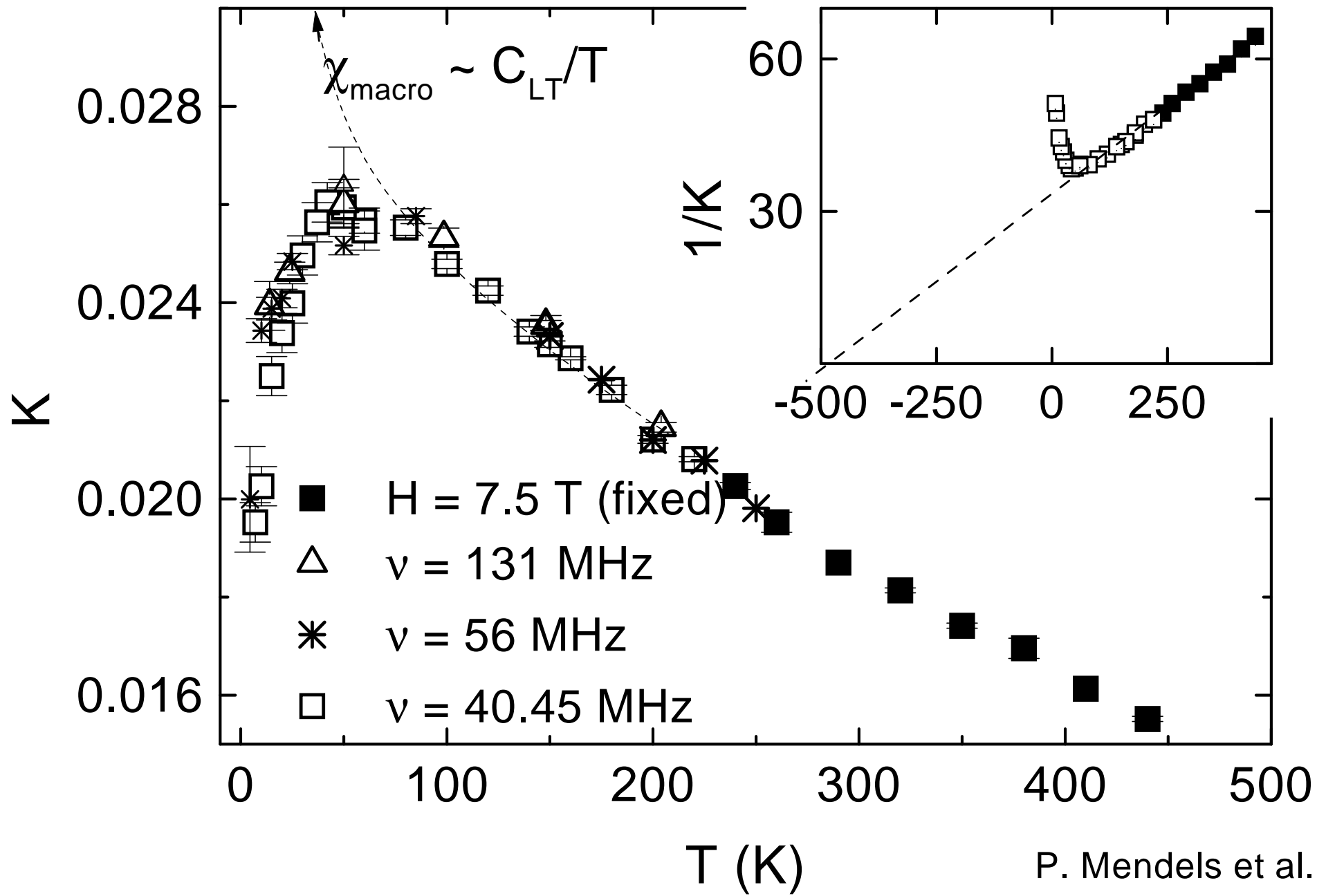
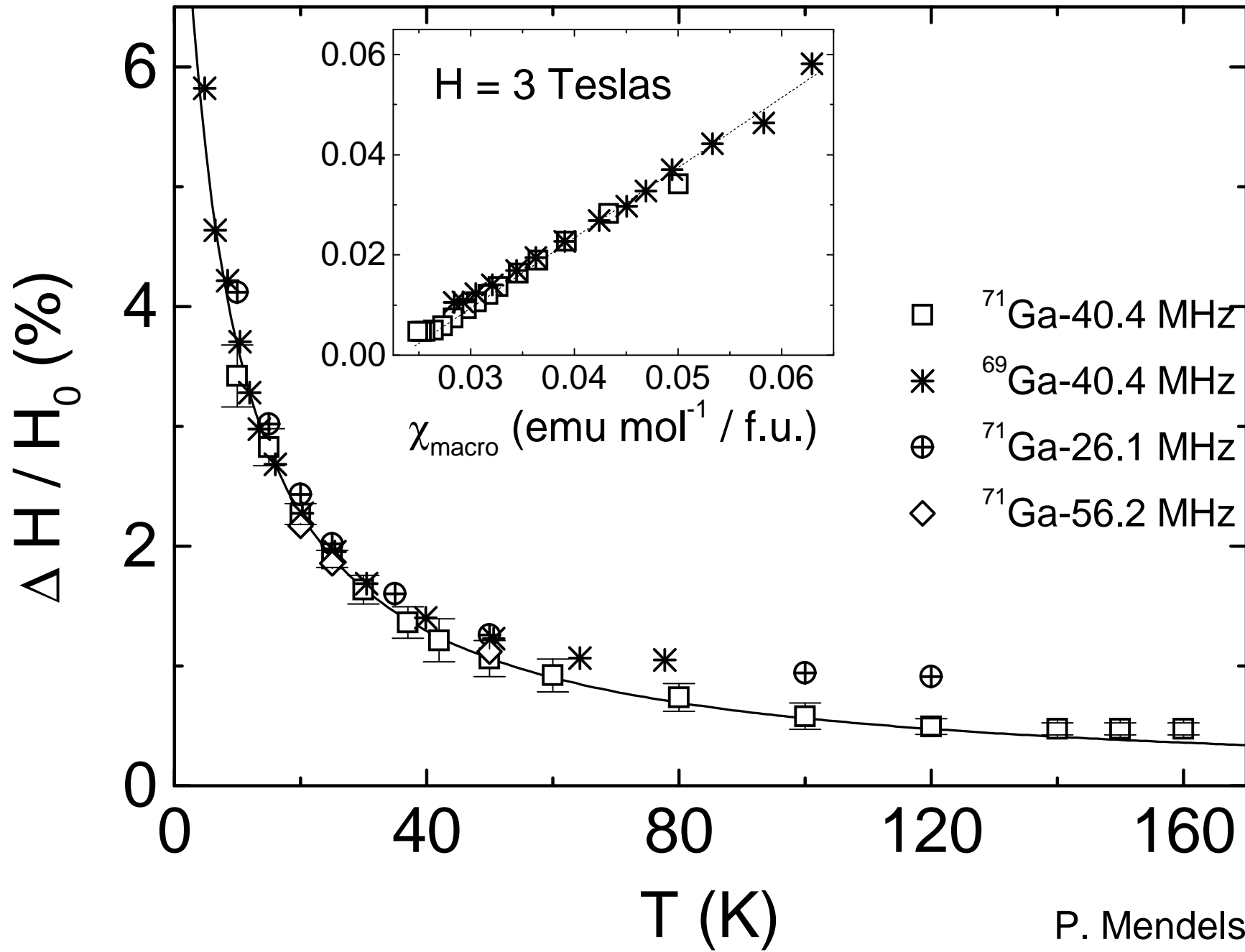
\end{document}